\documentclass[twocolumn,showpacs,preprintnumbers,superscriptaddress,amsmath,amssymb]{revtex4}

\usepackage{graphicx}
\usepackage{dcolumn}
\usepackage{bm}

\begin{document}


\title{Search for the proton decay $p\to K^{+}\overline{{\nu}}$ in the large liquid scintillator low energy neutrino astronomy detector LENA}

\author{T. Marrod\'an Undagoitia} 
\altaffiliation{Corresponding author. Fax: +49 89 289 12680,
Telephone Number: +49 89 289 12328.} 
\email{tmarroda@ph.tum.de}
\author{F. von Feilitzsch}
\author{M. G\"oger-Neff} 
\affiliation
{Physik-Department E15, Technische Universit\"at M\"unchen,\\
James-Franck-Str., 85748 Garching, Germany} 
\author{C. Grieb}
\affiliation{Virginia Polytechnic Institute and State University, \\
Physics Department, Robeson Hall, Blacksburg,\\ VA 24061-0435, USA}
\author{K. A.  Hochmuth} 
\author{L. Oberauer}
\author{W. Potzel} 
\author{M. Wurm} 
\affiliation
{Physik-Department E15, Technische Universit\"at M\"unchen,\\
James-Franck-Str., 85748 Garching, Germany}

\begin{center}
Published in Phys. Rev. D {\bfseries 72}, 075014 (2005) 
\end{center}

\begin{abstract}
The LENA ({\bfseries L}ow {\bfseries E}nergy {\bfseries N}eutrino
{\bfseries A}stronomy) detector is proposed to be a large-volume
liquid-scintillator device which will be highly suitable for the
investigation of a variety of topics in astrophysics, geophysics and
particle physics. In this paper, the potential of such a detector
concerning the search for proton decay in the SUSY favored decay
channel $p\to K^{+}\overline{{\nu}}$ is investigated. Based on Geant4,
Monte Carlo simulations of the proton decay in the LENA detector as
well as of the background radiation in the detection energy windows
have been developed.  From these simulations an efficiency of 65\% for
the detection of a possible proton decay has been determined. Within
ten years of measuring time a lower limit for the proton lifetime,
concerning the decay channel investigated, of
$\tau>4\cdot10^{34}~\textrm{y}$ could be reached.
\end{abstract}


\maketitle

\section{\label{DetectorDescription}Detector Description}

The LENA ({\bfseries L}ow {\bfseries E}nergy {\bfseries N}eutrino
{\bfseries A}stronomy) detector is planned~\cite{Ob05} to have a
cylindrical shape with about $100$~m length and 30~m diameter. An
inside part of 13~m radius will contain approximately 50~kt of liquid
scintillator while the outside part will be filled with water to act
as a muon veto. A fiducial volume for proton decay will be defined
having a radius of 12~m. Covering about 30$\%$ of the surface, 12\,000
photomultipliers of 50~cm diameter each will collect the light
produced by the scintillator. PXE (phenyl-o-xylylethane) is foreseen
as scintillator solvent because of its high light yield and its save
handling procedures. The optical properties of a liquid scintillator
based on PXE have been investigated in the Counting Test Facility
(CTF) for BOREXINO at the Gran Sasso underground
laboratory~\cite{CTF}. A yield of 372~$\pm$~8 photoelectrons per MeV
(pe/MeV) have been measured in this experiment with an optical
coverage of 20$\%$. The attenuation length of $\sim$~3~m (at 430~nm)
was substantially increased to $\sim$~12~m purging the liquid in a
weak acidic alumina column~\cite{CTF}. With these values an expected
photoelectron yield of $\sim$~120~pe/MeV can be estimated for events
in the center of the LENA detector. Currently the optical properties
of mixtures of PXE and derivatives of mineral oils are under
investigation~\cite{PXE}.

In addition to astrophysical processes such as supernovae neutrinos,
relic supernovae neutrinos and solar neutrinos also geophysical
aspects and research in particle physics, e.g. long baseline
experiments and the search for the proton decay will play a major
role. Possible locations for this detector, currently under
discussion, are an underground mine in the center of Finland
(Pyh\"asalmi, CUPP: Center of Underground Physics in Pyh\"asalmi) or
under water, close to the coast at Pylos in Greece. Both sites have a
shielding of $\sim$~4\,000 meters water equivalent (m.w.e.) and are
far away from nuclear power plants, which cause the main contribution
to the $\overline{{\nu}}_{e}$ background in the measurement of relic
supernovae neutrinos. 

In this letter the sensitivity of LENA for the search of the proton
decay $p\to K^{+}\overline{{\nu}}$ is investigated. The potential of 
a large liquid scintillator detector for proton decay detection has
already been proposed by R. Svoboda~\cite{Svoboda}.

\section{\label{TheoreticalPredictions}Theoretical Predictions}

In the Grand Unified Theories (GUTs), the decay of the proton is
predicted, leading to a nonconservation of the baryon (B) and lepton
(L) numbers.  A measurement of such an event could be a probe of the
validity of those theories and further evidence for the existence of
physics beyond the Standard Model.  In the minimal GUT
SU(5)~\cite{Georgi74}, the nucleon decays via exchange of gauge bosons
with GUT scale masses ($\sim10^{15}$~GeV). The dominant proton decay
mode predicted is $ p\to e^{+}\pi^{0}$ with a lifetime of $\tau\sim
10^{31}$~y.  The Super-Kamiokande Collaboration has set the current
lower limit for this channel to $\tau> 5.4\cdot10^{33}$~y at 90$\%$
confidence level (C.L.)~\cite{Hyperk}. This result together with the
fact that, in this theory, the three running coupling constants do not
meet in a single point give reasons to look also at other theories.

In Supersymmetric GUTs (SUSY), the nucleon decay is affected by
operators of dimension 4, 5 and 6. They involve B- and L-violating
processes but conserve B-L. For the proton decay mode $p\to
e^{+}\pi^{0}$ a lifetime of $\tau\sim 5\cdot10^{35\pm1}$~y has been
calculated. This rather long lifetime results from the large value of
the unification mass, $M_{G}$, given by those theories. For the decay
channel $p\to K^{+}\overline{{\nu}}$ ($\overline{{\nu}}=\sum_{\{
i=e,\mu,\tau\}}\overline{{\nu_{i}}}$) a range for the lifetime
$\tau\sim(0.33-3)\cdot10^{34}$~y~\cite{Raby02} has been
predicted. Also previous studies on SUSY theory have found the same
decay channel $p\to K^{+}\overline{{\nu}}$ to be the dominant one with
a lifetime range of $\tau\sim(10^{30}-10^{35}$~y)~\cite{Babu98}. The
kaon produced in the dominant SUSY proton decay channel is `invisible'
in water Cherenkov detectors because its momentum is below the
threshold for producing Cherenkov light
($T_{th}(\textrm{K}^{+})=253$~MeV). For this reason, the
Super-Kamiokande experiment could only perform an indirect measurement
with efficiencies of 4.4\% and 6.5\% corresponding to two different
methods~\cite{SK99}. The lower limit that they reach at 90$\%$ C.L. is
$\tau>2.3\cdot10^{33}$~y~\cite{Super05}.

Supergravity theories provide also predictions for the possible decay
channels of the proton. The dominant mode here is
$p\to\pi^{+}\overline{{\nu}}$ with a branching ratio of
65.7$\%$. However, again the proton decay mode $p\to
K^{+}\overline{{\nu}}$ is predicted with a branching ratio of
33.5$\%$~\cite{Ar85}.

In a scintillation detector like LENA, the $\textrm{K}^{+}$ particles
from proton decay events can be observed directly. These signals
together with the delayed signals from the decay products provide a
very distinct signature for this decay mode.

\section{\label{DetectionMechanism}Detection Mechanism}

Within the detector volume of LENA, about 1.45$\cdot10^{34}$~protons,
both from C- and H-nuclei, are candidates for the decay. This number
has been calculated taking into account the chemical composition of
PXE ($\textrm{C}_{16}\textrm{H}_{18}$), its density
(0.985~g/$\textrm{cm}^{3}$) and its molecular weight
(210.3~a.m.u.). 

In the case of protons from H-nuclei ($0.23\cdot10^{34}$~protons in
the fiducial volume of LENA), the decay problem is simplified because
those protons decay at rest. Therefore, the proton decaying in the
channel $p\to K^{+}\overline{{\nu}}$ can be considered as a two-body
decay problem where $K^{+}$ and $\overline{{\nu}}$ always receive the
same energy. The energy corresponding to the mass of the proton,
$m_{p}=938.3$~MeV is thereby given to the decay products. Using
relativistic kinematics, it can be calculated that the
particles receive fixed kinetic energies, the neutrino 339~MeV and the
kaon 105~MeV. 

In the LENA detector, the kaon will cause a prompt mono-energetic
signal while the neutrino escapes without producing any detectable
signal. After $\tau( K^{+})=12.8$~ns, the kaon decays via $K^{+}
\to\mu^{+}{\nu_{\mu}}$ with a probability of 63.43~$\%$. In 90\% of
the times, the kaon decays at rest~\cite{SK99} and from this kaon
decay, a second mono-energetic signal arises, corresponding to the
$\mu^{+}$ with 152~MeV kinetic energy. Later, after
$\tau(\mu^{+})=2.2~\mu$s, also the muon will decay: $\mu^{+}\to
e^{+}{\nu_{e}}{\overline{\nu}_{\mu}}$, the $e^{+}$ will produce a
third long-delay signal. Since the last decay is a three body decay,
the $e^{+}$ does not have a fixed energy but a spectrum. With a
smaller probability, 21.13~$\%$, the kaon decays, alternatively, via
\mbox{$K^{+}\to\pi^{+} \pi^{0}$}. The $\pi^{+}$ deposits its kinetic
energy (108~MeV) in the detector. The $\pi^{0}$ decays almost
immediately, within $\tau(\pi^{0})= 8.4\cdot10^{-8}$~ns, into two
gammas with a total energy being equal to the sum of the kinetic
energy of the $\pi^{0}$ (111~MeV) and its rest mass
($m_{\pi^{0}}=135$~MeV). These gammas can be seen in the detector as
electromagnetic showers. Thus again there are two mono-energetic
signals, firstly the kaon and secondly the one corresponding to the
sum of the kinetic energy of the $\pi^{+}$ and the energy of the
gammas. Again, there will be a third long-delayed signal from the
succeeding decays ($\pi^{+}\to\mu^{+}\nu_{\mu}$ and $\mu^{+}\to
e^{+}{\nu_{e}}{\overline{\nu}_{\mu}}$). Only these two most likely
decay channels of the kaon are considered in the analysis.

For the protons of the C-nuclei ($1.22\cdot10^{34}$~protons in the
fiducial volume of the LENA detector), one has to consider further
nuclear effects on the proton decay. First of all, since the protons
are bound to the nucleus their effective mass will be reduced by the
binding energy. The proton energy available for distribution between
the decay products, will be $M'_{p}=M_{p}-E_{b}$, where $M'_{p}$ is
the modified proton mass, $M_{p}=938.3$~MeV is the rest mass of the
proton and $E_{b}$ is the nuclear binding energy. The value of the
binding energy $E_{b}$ can be derived from gaussian probability
functions which for C-nuclei~\cite{FermiMotion} are centered at 37~MeV
and 16~MeV for protons in s-state and p-states,
respectively. Secondly, in case of a proton decay in a nucleus, e.g.,
carbon, decay kinematics are different from the free proton due to the
Fermi motion of the proton. These Fermi momenta have been measured by
electron scattering on ${}^{12}$C~\cite{FermiMotion}. The maximum
momentum is about 250~MeV/c. 

To determine the kinetic energy change of the decay particles of the
proton, various decay processes have been
calculated~\cite{MyThesis}. First of all, the mass of the proton was
modified due to the binding energies for s- and p-states according to
the values mentioned above. For both, two extreme cases were
considered: the neutrino, after the decay, moving in the direction of
the original proton and the neutrino moving in the opposite direction,
the kaon, of course, always moving in the opposite direction of the
neutrino.  The momentum of the proton was chosen to be maximal,
corresponding to a Fermi momentum of $\sim$250~MeV/c. The limiting
values for the range of the kinetic energy of the kaon are:
$(25.1-198.8)$~MeV for the s-state and $(30.0-207.2)$~MeV for the
p-state.

\section{\label{Simulation}Simulation}

In order to quantitatively estimate the potential of the LENA detector
for measuring the proton lifetime, a Monte Carlo simulation for the
decay channel $p\to K^{+}\overline{{\nu}}$ has been performed. For
this purpose, the Geant4 simulation toolkit has been
used~\cite{Geant}. The event generator was programmed to produce
$\textrm{K}^{+}$ particles of 105~MeV in the center of the detector
volume and with random directions. The simulation was also performed
for proton decays which do not take place in the center to investigate
the dependence on the event position. 

Not only all default Geant4 physics lists were included but also
optical processes as scintillation, Cherenkov light production,
Rayleigh scattering and light absorption. For the last two processes
scattering and absorption lengths of $\lambda_{s}=60$~m and
$\lambda_{a}=12$~m, respectively, were assumed. With these numbers an
effective attenuation length of $\lambda=10$~m results from:

\begin{equation}\label{eq:Attenuation}
\lambda=\frac{\lambda_{s}\cdot\lambda_{a}}{\lambda_{s}+\lambda_{a}}
\end{equation}
This is a quite conservative approach compared to the measured value
of $\lambda=12$~m~\cite{CTF} and results in a light yield of
$\sim$~110~pe/MeV for an event in the center of the detector. The
dependence of the detection efficiency and the photoelectron yield on
scattering and absorption lengths will be discussed later (see Chapter
6, Table~\ref{Table2}). 

To determine the detector efficiency, a total of 20\,000 $p\to
K^{+}\overline{{\nu}}$ Monte Carlo events were generated. Within this
simulation, it was possible to verify that the kaons decay according
to the branching ratios for the different channels stated
in~\cite{PDG}. The amount of light emitted by a scintillating material
is not strictly proportional to the energy deposited by the ionizing
particle. In reality, it is a complex function of energy, the type of
particle and the specific ionization~\cite{Leo}. To take into account
the so called quenching effects, the semi-empirical Birk's
formula~\cite{Birk} has been introduced into the code:

\begin{equation}\label{eq:Birks}
\frac{\textrm{d}L}{\textrm{d}x} = \frac{A\frac{\textrm{d}E}{\textrm{d}x}}{1+kB\frac{\textrm{d}E}{\textrm{d}x}}
\end{equation}

The parameters of this formula are A: absolute scintillation
efficiency and kB: parameter relating the density of ionization
centers to $\frac{dE}{dx}$. Both are specific values for every
scintillator and are obtained from experimental data. For the
simulation, $A=10^{4}$~photons/MeV and $kB=1.5\cdot10^{-2}$~cm/MeV
have been taken from studies performed by the BOREXINO
Collaboration~\cite{Borex}. According to equation~\ref{eq:Birks}, the
program calculates in every step the deposited energy ($dE$) within
the distance ($dx$) travelled by the particle and derives the light
output ($\frac{dL}{dx}$).

Also the photomultipliers have been introduced in the simulation
assuming an efficiency of 17~$\%$ and a time jitter of $\sigma = 1$~ns
originating from the transit time of electrons through the
photomultipliers and from the statistical noise. These features are
known because this type of photomultiplier has already been used in
other experiments~\cite{Borex02}.

\section{\label{BackgroundRejection}Background Rejection}
\subsection{\label{MuonProduction}Muon Production}

The main background source in the energy range of the proton decay are
muon neutrinos $\nu_{\mu}$ produced by cosmic ray interactions in the
atmosphere. Atmospheric $\nu_{\mu}$ can interact with the scintillator
producing muons in the energy range where the search for the proton
decay is performed. The rate of these events in the relevant energy
range can be derived from the Super-Kamiokande
measurements~\cite{SK99}. From this data we find an event rate of:

\begin{equation}
\Gamma \sim 4.8\cdot 10^{-2} (\textrm{MeV}^{-1}\textrm{kt}^{-1}\textrm{y}^{-1})
\end{equation}
where the LENA energy window as well as the volume have to be
introduced in MeV and kt, respectively.

\begin{figure}[!hbp]
\includegraphics[width=0.5\textwidth]{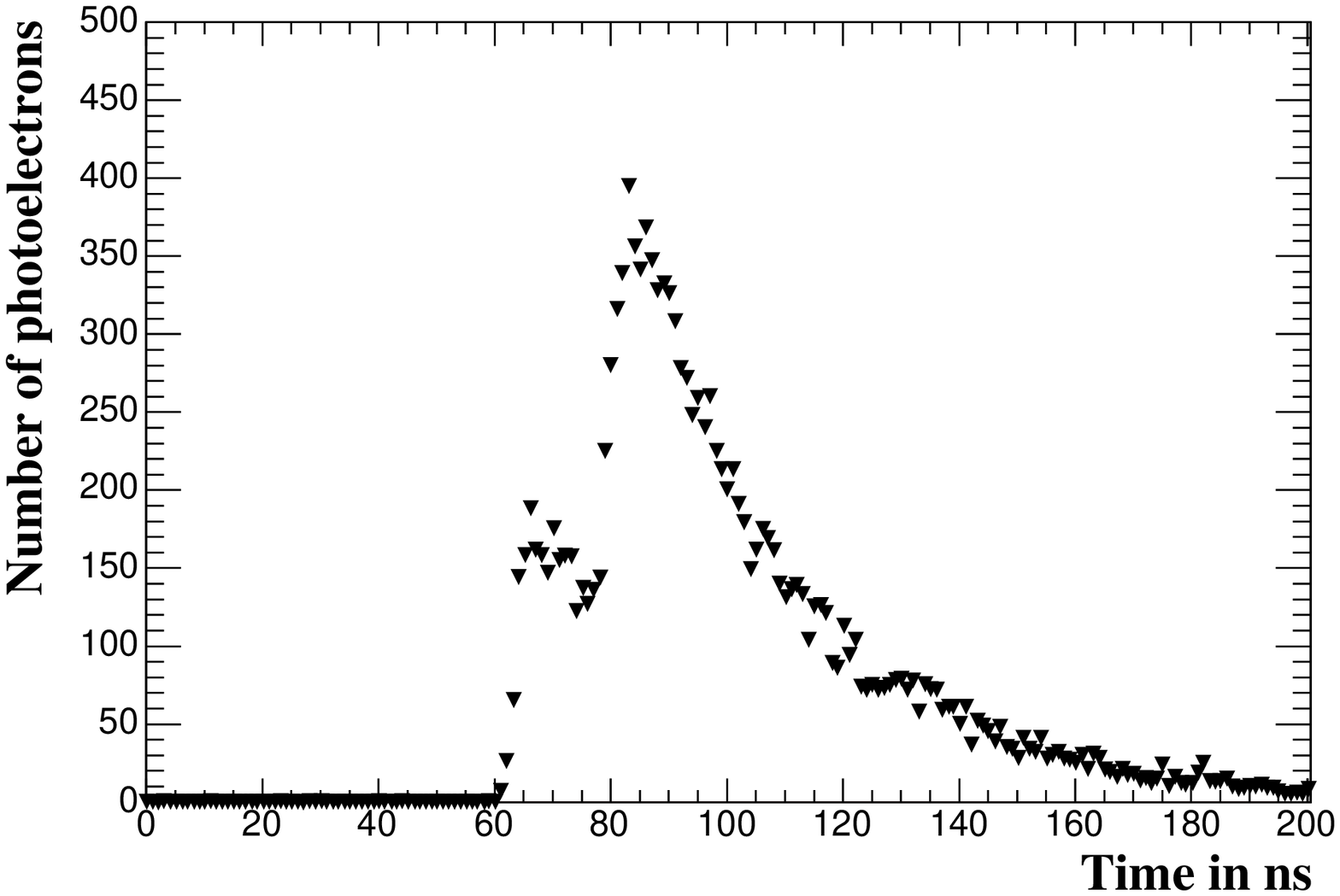}
\includegraphics[width=0.5\textwidth]{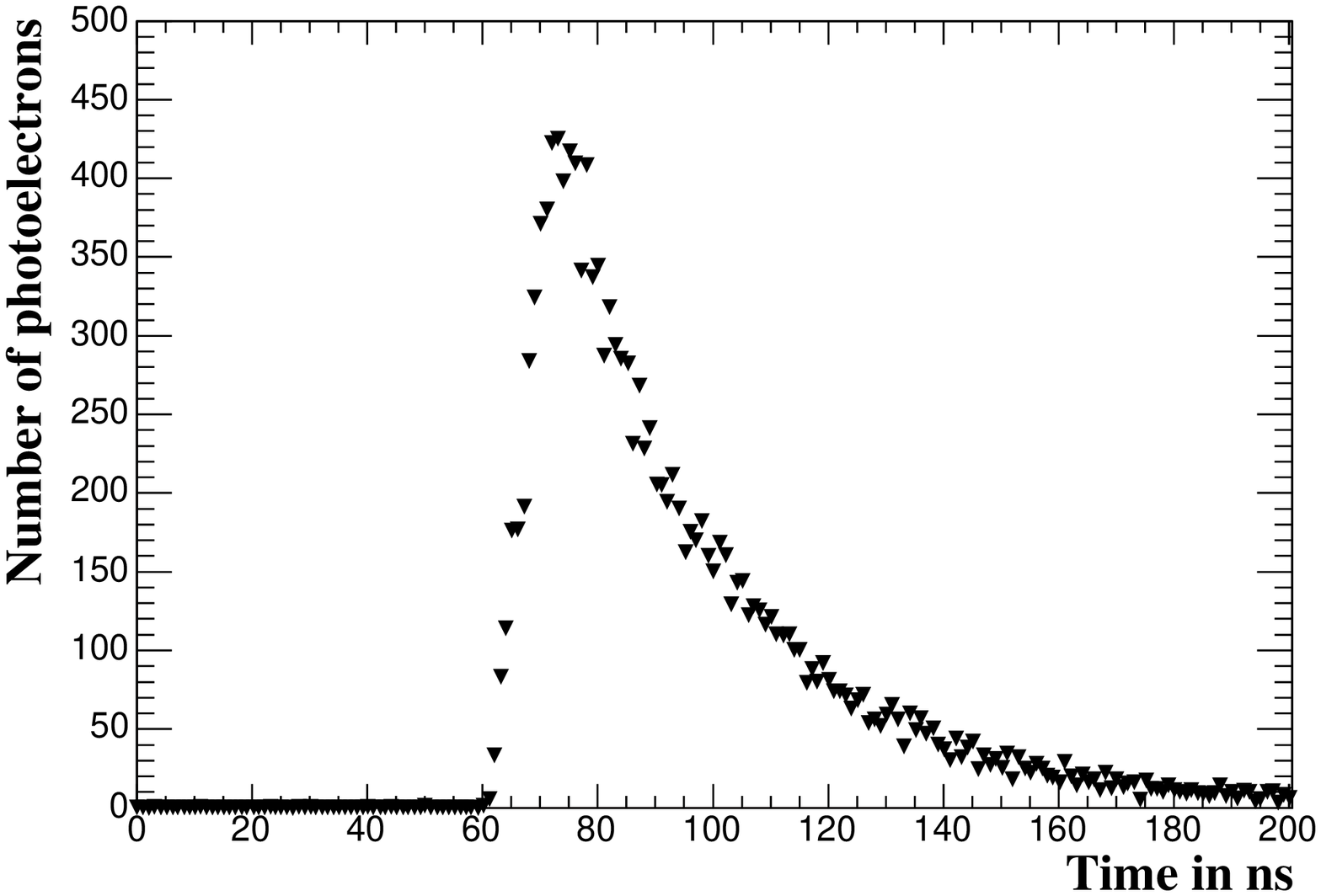}
\includegraphics[width=0.5\textwidth]{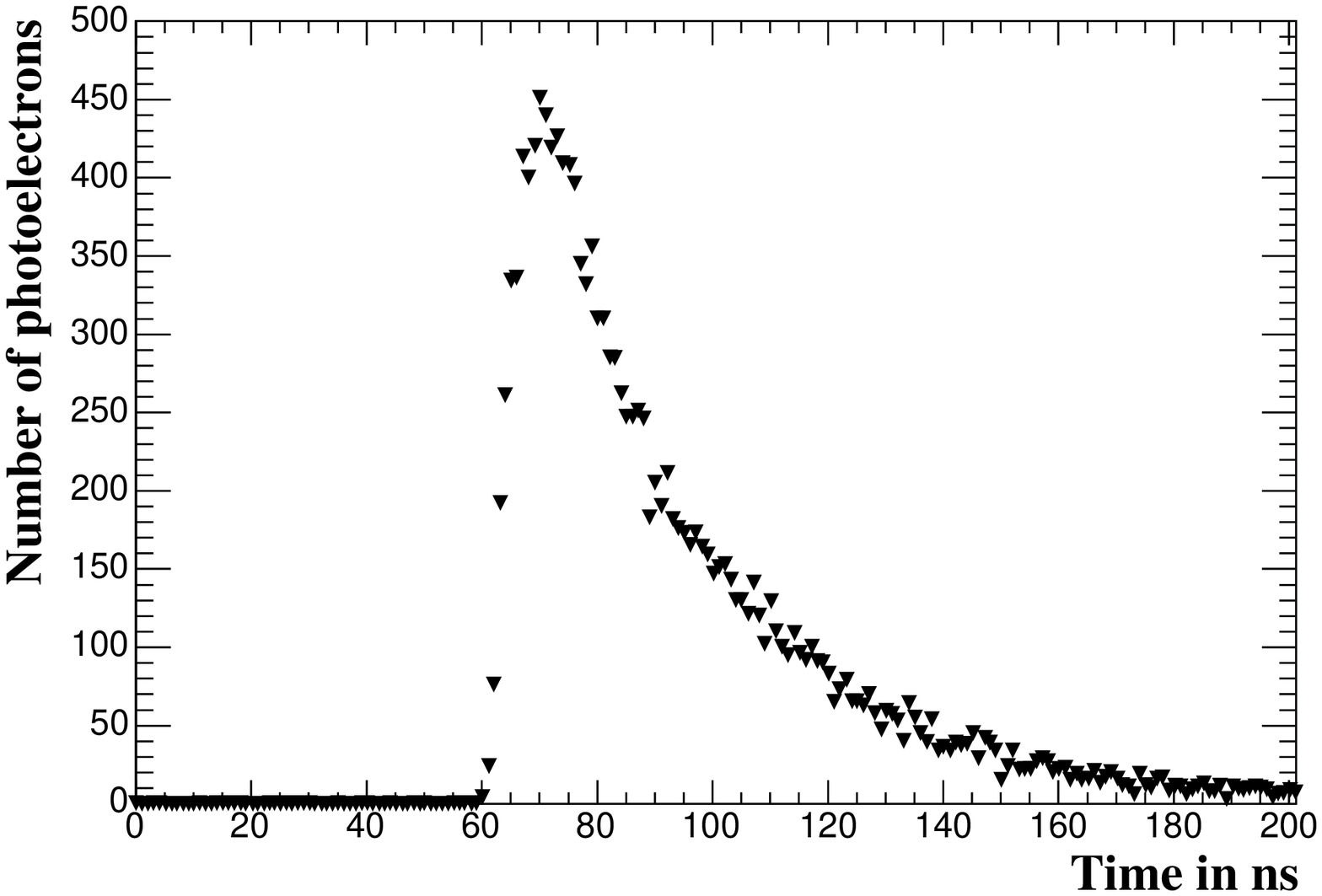}
\caption{Proton decay signal where the kaon decays after 18~ns (top) and after 5~ns (middle). The bottom panel shows a $\nu_{\mu}$ background signal.\label{Signals}} 
\end{figure}

In order to distinguish the real proton decay signals from muon
background events, a pulse shape analysis can be applied. From the
simulation it is known that the kaon deposits its energy in
1.2~ns. Then, as the kaon decays within a relatively short time
($\tau(\textrm{K}^+)=12.8$~ns), the signature from such an event will
consist of two shortly delayed signals. Examples of a proton decay
signal where the separation is clearly visible, where it is almost
hidden in the risetime and of a muon background signal can be seen in
figure~\ref{Signals}. The better the two signals can be separated, the
better the background suppression will be. In the cases where the kaon
decays at late times ($t>10$~ns), the two peaks can be easily
separated; in all other cases, only a shoulder within the risetime can
be seen.

The on-set time of these examples is 60~ns as the events are generated
at the center of the detector and the photons need this time to reach
the photomultipliers.

\subsubsection{\label{TimeCut}Time Cut}

\begin{figure}[!hbp]
\includegraphics[width=0.5\textwidth]{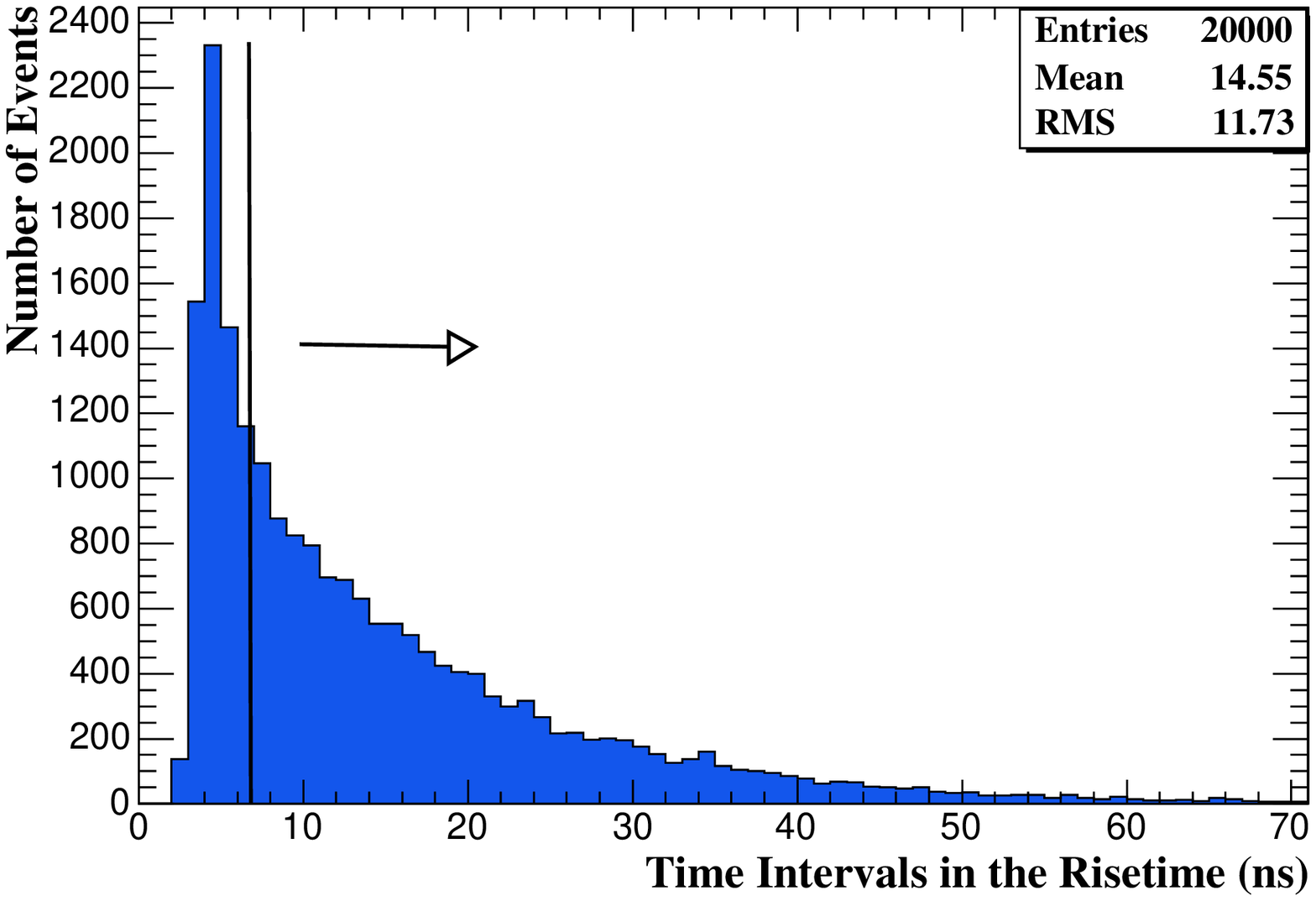}
\includegraphics[width=0.5\textwidth]{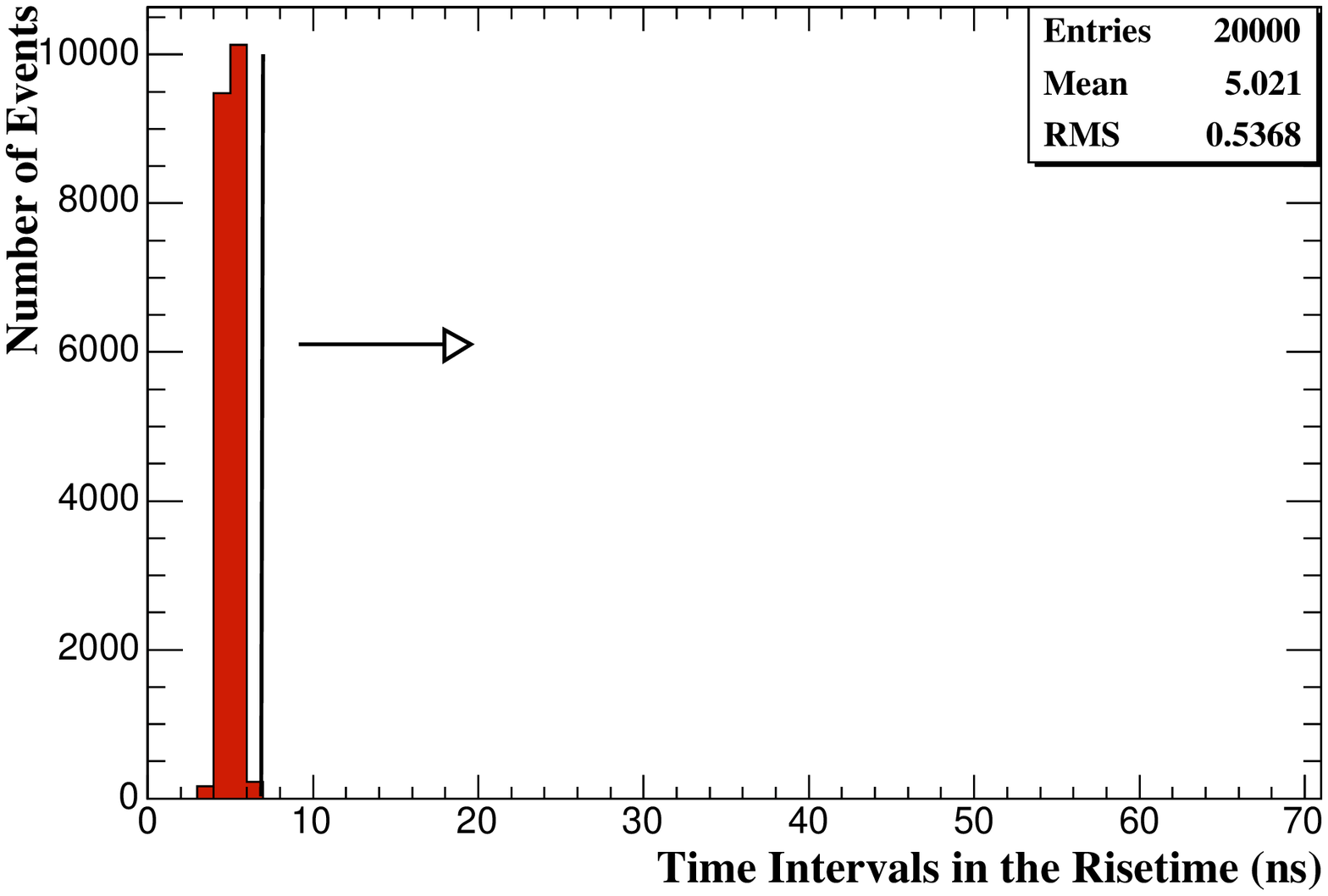}
\caption{Distribution of time intervals in the risetimes for proton decay (top) and muon background signals (bottom). A cut at 7~ns yields a reduction factor of $\sim2\cdot10^{4}$ on background events whereas the efficiency for observing a proton decay $p\to K^{+}\overline{{\nu}}$ is 65\%.\label{Intervals}}
\end{figure}

A first step towards a pulse shape analysis has been taken in order to
quantify the signal and background discrimination. Not only
20\,000~Monte Carlo proton decay events were generated but also
20\,000~muon background events for each kaon decay channel
analysed. For these simulations, the particles were generated in the
center of the detector and the light produced was followed through the
scintillator and the photomultipliers. This results in the number of
photoelectrons detected as a function of time. From the resulting
signal the maximum number of photoelectrons is taken and the 15~$\%$
and the 85~$\%$ values of the maximum signal height are
calculated. The difference of the two corresponding time values is
plotted in figure \ref{Intervals}. 

As an example, a cut at 7~ns is shown in this figure. Applying such a
cut, a background reduction by a factor of $\sim2\cdot10^{4}$ is
achieved. The background plot shown applies only to the channel
\mbox{ $p\to K^{+}\overline{{\nu}}$} and
$K^{+}\to\mu^{+}{\nu_{\mu}}$. A similar simulation has been performed
for the channel $p\to K^{+}\overline{{\nu}}$ and $K^{+} \to\pi^{+}
\pi^{0}$ resulting in an equivalent result where the same rejection
factor is obtained. An efficiency of \mbox{$\varepsilon_{T}= 0.65$}
results after this time cut because only 35\% of the signals are lost.

\subsubsection{Energy Cut}

\begin{figure}[!hbp]
\includegraphics[width=0.5\textwidth]{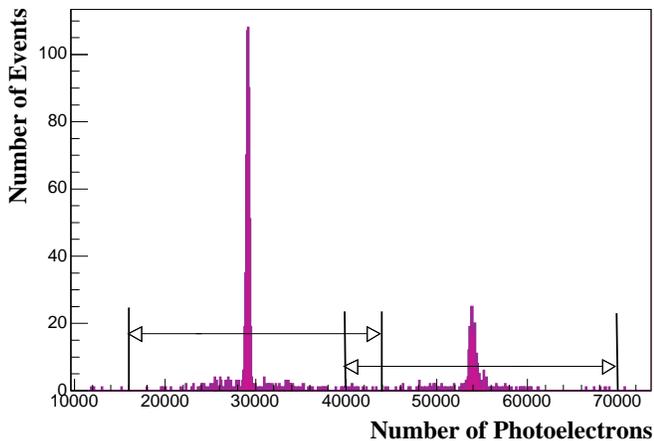} 
\caption{Distribution of the number of photoelectrons produced in the photomultipliers coming from free proton decay events at the center of the detector. The cuts performed for the two main decay channels of the kaon (at $\sim257$ and $\sim459$~MeV) are shown. It can be recognized that the energy windows taken overlap resulting in a total window of 55000~photoelectrons $\simeq$ 500~MeV.\label{Energy}} 
\end{figure}

In order to further improve the signal to background separation a cut
in the energy spectrum can be applied. Figure \ref{Energy}, shows the
distribution of the number of photoelectrons coming from free proton decay
events at the center of the detector as seen in the photomultipliers.

Two peaks can clearly be identified, corresponding to the two main
decay channels of the kaon. They are related to the energies of 257
and 459~MeV, which is the sum of the energies deposited by the kaon
and its decay products (see section Detection Mechanism). For each
peak a 300~MeV window has been set resulting in an efficiency of
$\varepsilon_{E}=0.995$ for this energy cut. The two energy windows
overlap resulting in a final window of 500~MeV. The signals outside
these energy windows correspond mainly to kaon decay channels not
considered in the analysis. There is also a small loss due to
inelastic hadronic interactions of the decay particles before they are
stopped.

\subsection{Hadron Production}

Atmospheric neutrinos can interact with the detector producing also
hadrons. The most probable of these reactions is the single pion
production~\cite{PionProd1}\cite{PionProd2}:

\begin{equation}\label{eq:PionProd}
{\nu}_{\mu}+p\to \mu^{-}+\pi^{+}+p'
\end{equation}

In such a reaction, the first signal comes from the sum of the
energies deposited by the $\mu^{-}$ and the $\pi^{+}$. Later, after
$\tau_{\pi^{+}}=26$~ns, the $\pi^{+}$ decays into $\mu^{+}$ and one
neutrino producing a second short-delay signal. However, the $\pi^{+}$
has such a small mass (139.6~MeV) that its decay muon receives only
about 20~MeV kinetic energy. The signal produced by the $\mu^{+}$ is
small. Therefore, those events do not have the same signature as
the proton decay signal. Moreover, since the muon signal is so small
it is usually hidden behind the above mentioned first signal. 

Neutrinos of higher energies can interact producing strange particles,
thus also kaon production has to be considered. According to the
MINERvA proposal~\cite{Minerva}, the reactions of this type with
notable probability are:

\begin{equation}\label{eq:pKProd}
{\nu}_{\mu}+p\to \mu^{-}+K^{+}+p
\end{equation}
\begin{equation}\label{eq:KLambProd}
{\nu}_{\mu}+n\to \mu^{-}+K^{+}+\Lambda^{0}
\end{equation}
\begin{equation}\label{eq:KLambPiProd}
{\nu}_{\mu}+n\to \mu^{-}+K^{+}+\Lambda^{0}+\pi^{0}
\end{equation}

\begin{figure}[!hbp]
\includegraphics[width=0.5\textwidth]{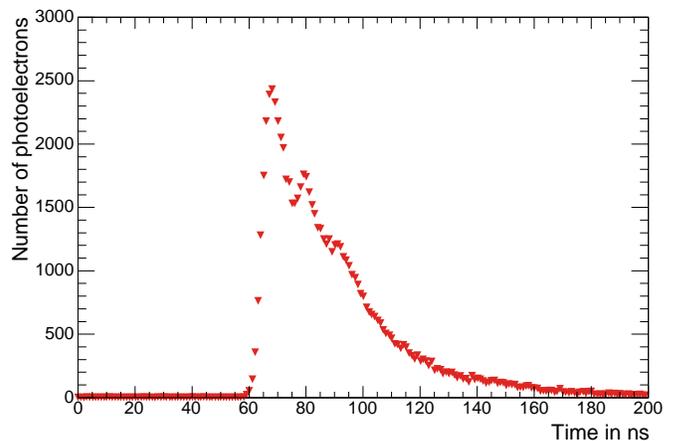}

\caption{Background signal for proton decay events. The first peak
corresponds to the signals of the $\mu^{-}$, $K^{+}$ and
$\Lambda^{0}$. At later times, two other peaks appear, one is caused
by the $\mu^{+}$ that comes from the decay of the $K^{+}$ and the
other a $\mu^{-}$ from $\Lambda^{0}\to p+\pi^{-}$ and
$\pi^{-}\to\mu^{-}\overline{\nu}_{\mu}$.\label{BgMuKLambda}}
\end{figure}

In equations~\ref{eq:KLambProd} and~\ref{eq:KLambPiProd}, to conserve
the strangeness number the kaon is produced together with a
$\Lambda^{0}$ baryon. This $\Lambda^{0}$ particle decays after
$\tau_{\Lambda^{0}}=0.26$~ns mainly via two channels $\Lambda^{0}\to
p+\pi^{-}$ (63.9\%) or $\Lambda^{0}\to n+\pi^{0}$ (35.8\%). Because of
this short lifetime, the p, $\pi^{-}$ or n, $\pi^{0}$ cause a prompt
signal together with the $\mu^{-}$, the $K^{+}$ and the $\pi^{0}$ in
the case of equation~\ref{eq:KLambPiProd}. The prompt signal in these
two reactions is bigger than the signal of the $\mu^{+}$ from the kaon
decay (or $\pi^{+}$, $\pi^{0}$ if the kaon decays within the
second-probable channel). For this reason, the signature of the
background signal can be distinguished from proton decay
events. Figure \ref{BgMuKLambda} shows an example of
equation~\ref{eq:KLambProd}. First a big peak can be seen
corresponding to the signals of the $\mu^{-}$, $K^{+}$ and
$\Lambda^{0}$. After that, two other peaks appear, one is due to the
$\mu^{+}$ that originates from the decay of the $K^{+}$ and the other
a $\mu^{-}$ from $\Lambda^{0}\to p+\pi^{-}$ and
$\pi^{-}\to\mu^{-}\overline{\nu}_{\mu}$.

Reaction~\ref{eq:pKProd} can be responsible for a potential background
for proton decay in the LENA detector. For this case, only neutrinos
with energies between 650 and 900~MeV can produce a signal with a
signature similar to that of the proton decay. This neutrino energy
window has been determined considering the energy window of proton
decay events relevant for the analysis ($150 - 650$~MeV) and the
energies deposited by the particles in the possible background
events. The atmospheric neutrino flux has a maximum around 100~MeV and
then decreases exponentially to higher energies. Assuming an energy
dependence of $\phi_{\nu}\sim E_{\nu}^{-2.7}$ for the neutrino flux
and a linear dependence for the cross-section, 10\% of the total
atmospheric neutrino flux occurs within the energy window considered
in our analysis. Within this window a rate of 0.8 events per year
($y^{-1}$) caused by such neutrino reactions is predicted. To
distinguish these background reactions from proton decay events the
number of delayed electrons produced will be taken into account. For
proton decay in the channel considered, always one and only one
electron is produced by the decay of the kaon. In the background
reaction always two electrons are present, one from the $\mu^{-}$
decay and another from the $K^{+}$ decay chain. This fact considerably
reduces the background contribution. Only in those cases where the
energy of one of the background electrons is very small (E $<$
0.5~MeV) or when the signal of these electrons is hidden under that of
the parent particle, the process exhibits only one electron. Events
with electrons being hidden within the previous signal happen in 4\%
of the cases. From 200 muons simulated none of them had a decay
electron with an energy smaller than 0.5~MeV. To estimate the number
of background events of this type, the rate predicted has to be
multiplied by this 0.04 and then, by two because electrons both from
the $\mu^{-}$ and the $K^{+}$ decay chain can be hidden within the
previous signal. Thus, the number of events of the type of
reaction~\ref{eq:pKProd} expected in the LENA detector is
0.064~$y^{-1}$. 

Other reactions where hadrons are produced are possible but all of
them have lower rates and can also be distinguished from a proton
decay signal by applying the arguments used for
reactions~\ref{eq:KLambProd} and~\ref{eq:KLambPiProd}.

\subsection{Neutron Production by Cosmic Muons}

Also very fast neutrons, generated by spallation processes of cosmic
muons outside the detector, may reach the scintillator causing
background. These neutrons are slowed down through scattering
processes and the recoil protons may produce a background signal in
the detector. However, only neutrons produced by muons passing close
to the detector but not through the veto, have to be taken into
account. In addition, these signals can be tagged as the neutrons are
captured by hydrogen in the scintillator after $\sim$~200~$\mu$s,
producing an observable 2.2~MeV $\gamma$-ray.

\section{\label{ProtonDecaySensitivity}Proton Decay Sensitivity}

Once the time cut and the energy cut have been applied, one can
calculate the final efficiency $\varepsilon = \varepsilon_{E}\cdot
\varepsilon_{T}$ of the proton decay detection in the channel
considered. In table~\ref{Table1} the efficiencies for events at
different points of the detector are presented as well as the
background rates for the same positions. As the detector is
symmetrical along the cylinder axis the points considered are in the
middle of the cylinder and at different radial distances. No
dependence of the efficiency on the event position was found within
the statistical error. This statistical error mainly comes from the
time cut as the one from the energy cut is negligible. The background
rates presented have been obtained by multiplying the muon background
rate in the LENA detector (1190.4~$\textrm{y}^{-1}$) derived from the
Super-Kamiokande experiment~\cite{SK99} with the background
suppression from the time cut ($5\cdot10^{-5}$). The contribution of
the kaon production background (0.064~$\textrm{y}^{-1}$) has been
added to the previous value.

\begin{table}
\caption{\label{Table1}For 1000 events produced at each radial position $R$ in the detector, the efficiency in the time-cut $\varepsilon_{T}$, the efficiency in the energy-cut $\varepsilon_{E}$, the total efficiency $\varepsilon$ including the statistical error, and the background rate $B$ are given.}
\begin{ruledtabular}
\begin{tabular}{|c|c|c|c|c|}
\hline
$R$ (m) & $\varepsilon_{T}$ & $\varepsilon_{E}$ & $\varepsilon$ ($\pm0.04$) & $B$ ($\textrm{y}^{-1}$) \\
\hline 
Center & 0.649  & 0.995  & 0.65 & 0.11 \\
\hline 
3.0  & 0.675  & 0.994  & 0.67 & 0.11 \\
\hline 
6.0  & 0.650  & 0.994  & 0.65 & 0.11 \\
\hline
9.0  & 0.679  & 0.995  & 0.68 & 0.11 \\
\hline
11.5 & 0.666  & 0.995  & 0.66 & 0.11 \\
\hline
\end{tabular}
\end{ruledtabular}
\end{table}

In table~\ref{Table2} the efficiencies for proton decay detection and
the photoelectron yield are given as function of different
attenuation, absorption and scattering lengths. For the results
presented, 1000 proton decay and 1000 background events were simulated
for each combination of $\lambda_{a}$, $\lambda_{s}$ and
$\lambda$. All events were programmed to take place at the center of
the detector. The last column of the table shows the time (see
figure~\ref{Intervals}) at which the cut can be performed such that
all background signals are rejected. In the upper part of the table,
three attenuation lengths $\lambda$ are presented keeping the
contributions of absorption length and scattering length equal,
$\lambda_{a}=\lambda_{s}$. As the attenuation length increases the
photoelectron yield Y increases and the cut in the time intervals can
be performed earlier. In the bottom part, the attenuation length has
always the same value $\lambda=10$~m and different values for the
absorption and scattering lengths are given satisfying equation
(1). The photoelectron yield increases as the absorption length
increases because more optical photons reach the photomultipliers and
therefore more photoelectrons are produced. As the scattering length
increases, the cut in the time intervals can be performed earlier
because the time information is better.

\begin{table}
\caption{\label{Table2}Efficiencies $\varepsilon$ for the detection of
proton decay and photoelectron yield as function of attenuation,
absorption and scattering lengths. The last column shows the time (see
figure~\ref{Intervals}) at which the cut can be performed such that
all background signals are rejected. Upper part: Different attenuation
lengths for $\lambda_{a}$ = $\lambda_{s}$. Bottom part: Attenuation
length $\lambda$=10~m for different combinations of $\lambda_{a}$ and
$\lambda_{s}$.}
\begin{ruledtabular}
\begin{tabular}{|c|c|c|c|c|c|}
\hline
$\lambda$ (m) & $\lambda_{a}$ (m)& $\lambda_{s}$ (m)& $\varepsilon$ & Y (pe/MeV) & Cut (ns)\\
\hline 
5  & 10 & 10 & 0.56 & 58 & 10\\
\hline 
7  & 14 & 14 & 0.65 & 116 & 8\\
\hline 
9  & 18 & 18 & 0.67 & 161 & 7\\
\hline \hline
10 & 12 & 60 & 0.65 & 110 & 7\\
\hline
10 & 15 & 30 & 0.69 & 145 & 7\\
\hline
10 & 20 & 20 & 0.66 & 180 & 7\\
\hline
10 & 30 & 15 & 0.63 & 230 & 8\\
\hline
10 & 60 & 12 & 0.62 & 303 & 9\\
\hline
\end{tabular}
\end{ruledtabular}
\end{table}

Table~\ref{Table2} shows that in all cases investigated, for $\lambda>
7$~m the efficiencies $\varepsilon> 0.62$. This is also true for
$\lambda_{a}=12$~m and $\lambda_{s}=60$~m, which we used for deriving
the results presented earlier in figures~\ref{Signals}
and~\ref{Intervals}. It demonstrates that the efficiency reachable is
about tenfold enhanced compared to the efficiency reached in
Super-Kamiokande for the same decay channel~\cite{SK99}.

The activity for the proton decay is given by the expression:

\begin{equation}
 A=\varepsilon N_{p}t_{m} / \tau 
\end{equation}
where $\varepsilon=0.65$ is the efficiency already explained;
$N_{p}=1.45\cdot 10^{34}$ is the number of protons in the detector;
$t_{m}$ is the measuring time and $\tau$ is the lifetime of the
proton.

For the current proton lifetime limit for the channel considered
($\tau=2.3\cdot10^{33}$~y)~\cite{Super05}, about 40.7 proton decay
events would be observed in LENA after a measuring time of ten years
with about 1.1 background events. If no signal is seen in the detector
within this ten years, the lower limit for the lifetime of the proton
will be placed at $\tau>4\cdot10^{34}~\textrm{y}$ at $90\%$ C.L. If
one candidate is observed, the lower limit will be reduced to
$\tau>3\cdot10^{34}~\textrm{y}$ at $90\%$ C.L. and the probability of
this event being background would be 32\%.

\section{\label{Conclusions}Conclusions}

Within the simulations performed, an efficiency of $\sim$~65\% for the
search for proton decay in the LENA detector has been determined. A
lower limit for the proton lifetime of $\tau>4\cdot10^{34}~\textrm{y}$
(at $90\%$ C.L.) can be reached if no proton decay event is measured
within ten years.

The success in reaching this limit is based on the distinct pulse
shape of the proton decay mode $p\to K^{+}\overline{{\nu}}$ which is
clearly observable in the LENA detector. Only a minor energy cut was
used and hence the efficiency for observing the proton decay including
free and bound protons is high. Since the values predicted by the
favoured theories for the proton decay in this channel are of the
order of the value resulting from our simulation, it is obvious that
LENA measurements would have a deep impact on the proton decay
research field. In order to further improve the results of the present
simulation pulse shape analysis of higher precision will be performed.

\section*{\label{Acknowlegments}Acknowlegments}

We want to thank Prof. M. Lindner for the valuable discussions related
to theory, D. D'Angelo, Dr. C. Lendvai and Dr. L. Niedermeier for
detailed advice based on their experience from the BOREXINO experiment
and Q. Weitzel for continuous help. This work has been supported by
funds of the Maier-Leibnitz-Laboratorium (Garching) and by the
Deutsche Forschungsgemeinschaft DFG (Sonderforschungsbereich 375).

\newpage 
\bibliography{PRDSearchProtonDecayKNu2}

\end{document}